\documentstyle[12pt,epsfig,rotating]{article}     
\begin{document}

%------------------------------------------------------------------------
% new definitions, abreviations, etc
%------------------------------------------------------------------------

\def\beq             {\begin{equation}}
\def\eeq             {\end{equation}}
\def\beqd            {\begin{displaymath}}
\def\eeqd            {\end{displaymath}}
\def\baa             {\begin{array}}
\def\eaa             {\end{array}}
\def\beqaa           {\begin{eqnarray}}
\def\eeqaa           {\end{eqnarray}}
\def\beqaad          {\begin{eqnarray*}}
\def\eeqaad          {\end{eqnarray*}}
\def\btabu           {\begin{tabular}}
\def\etabu           {\end{tabular}}
\def\bfig            {\begin{figure}}
\def\efig            {\end{figure}}
\def\bce             {\begin{center}}
\def\ece             {\end{center}}

\def\ind             {\indent}
\def\noi             {\noindent}
\def\nn              {\nonumber}

\def\lbr             {\lbrack}
\def\rbr             {\rbrack}
\def\ti              {\tilde}
\def\q               {\bar}

\def\a               {\alpha}
\def\b               {\beta}
\def\d               {\delta}
\def\g               {\gamma}
\def\G               {\Gamma}
\def\l               {\lambda}
\def\t               {\theta}
\def\s               {\sigma}
\def\x               {\chi}

\def\de              {\partial}
\def\demu            {\partial_{\mu}}
\def\delr            {\!\stackrel{\leftrightarrow}{\partial^\mu}\!}

\def\sq              {\ti q}
\def\sqi             {\ti q_{i}^{}}
\def\sqL             {\ti q_{L}^{}}
\def\sqR             {\ti q_{R}^{}}
\def\sqe             {\ti q_{1}^{}}
\def\sqz             {\ti q_{2}^{}}
\def\sqp             {\ti q^{\prime}}
\def\qp              {q^{\prime}}

\def\su              {\ti u}
\def\sd              {\ti d}

\def\st              {\ti t}
\def\sb              {\ti b}

\def\ch              {\ti \x^\pm}
\def\chp             {\ti \x^+}
\def\chm             {\ti \x^-}
\def\nt              {\ti \x^0}

\def\Hk              {H_{\!k}^{}}
\def\sg              {{\ti g}}

\def\mH              {m_{\!H}}
\def\msg             {m_{\sg}}

\newcommand{\msq}[1]   {m_{\sq_{#1}}}
\newcommand{\mst}[1]   {m_{\st_{#1}}}
\newcommand{\msb}[1]   {m_{\sb_{#1}}}
\newcommand{\mch}[1]   {m_{\ti \x^\pm_{#1}}}
\newcommand{\mnt}[1]   {m_{\ti \x^0_{#1}}}

\def\hg              {^{(g)}}
\def\hsg             {^{(\sg )}}
\def\hsq             {^{(\sq )}}

\def\tW              {\t_W}
\def\sW              {\sin^2\t_W}

\def\tsq             {\t_{\sq}}
\def\sth             {\sin\t}
\def\cth             {\cos\t}
\def\sthq            {\sin^2\t}
\def\cthq            {\cos^2\t}
\def\dth             {\d\theta}

\def\kr              {^{\dagger}}

\def\lag             {{\cal L}}
\def\lum             {{\cal L}}
\def\R               {{\cal R}}
\def\Oas             {{\cal O}(\alpha_{s})}
\def\Emiss           {E\llap/}

\def\BR              {\rm BR}
\def\gev             {\:{\rm GeV}}
\def\pb              {${\rm pb}^{-1}$}

\def\rzw             {\sqrt{2}}

\def\ra              {\rightarrow}

\def\leerz           {\hspace{1cm}\\}

% Def. fuer kleine \fracs:
\newcommand{\smaf}[2] {{\textstyle \frac{#1}{#2} }}
\def\onesq            {{\textstyle \frac{1}{\sqrt{2}} }} 
\def\onehf            {{\textstyle \frac{1}{2} }} 
\def\oneth            {{\textstyle \frac{1}{3} }}  
\def\twoth            {{\textstyle \frac{2}{3} }}
\def\onefo            {{\textstyle \frac{1}{4} }}
\def\forth            {{\textstyle \frac{4}{3} }}

% Def. fuer Struktur {\cot\b, \tan\b}
\newcommand{\poss}[2] { \left\{ \!\! 
   {\footnotesize\begin{array}{r} #1 \\ #2 \end{array}} \!\!\right\} }

% Def. fuer groesser-ungefaehr:
\newcommand{\gsim}{\;\raisebox{-0.9ex}
           {$\textstyle\stackrel{\textstyle >}{\sim}$}\;}
% Def. fuer kleiner-ungefaehr:
\newcommand{\lsim}{\;\raisebox{-0.9ex}{$\textstyle\stackrel{\textstyle<}
           {\sim}$}\;}

% Def.s fuer Querverweise
\newcommand{\eq}[1]  {\mbox{(\ref{eq:#1})}}
\newcommand{\fig}[1] {\mbox{Fig.~\ref{fig:#1}}}

%------------------------------------------------------------------------
% title
%-----------------------------------------------------------------------

\begin{flushright}
  hep-ph/9806299\\
\end{flushright}

\bce

\vspace*{5mm}

{\Large {\bf SUSY-QCD corrections to\\[2mm] 
             stop and sbottom decays into Higgs bosons }} \\  

\vspace{12mm}

{\sc A. Bartl$^{\small 1}$,
     H. Eberl$^{\small 2}$,
     K. Hidaka$^{\small 3}$, } \\[2mm]
{\sc S. Kraml$^{\small 2}$,
     W. Majerotto$^{\small 2}$,
     W. Porod$^{\small 1}$,
     Y. Yamada$^{\small 4}$}

\vspace{10mm}

{\em (1) Institut f\"ur Theoretische Physik, Universit\"at Wien,
         A--1090 Vienna, Austria} \\[1mm]
{\em (2) Institut f\"ur Hochenergiephysik,
         \"Osterreichische Akademie der Wissenschaften, \\
         A--1050 Vienna, Austria}\\[1mm]
{\em (3) Department of Physics, Tokyo Gakugei University, Koganei,
         Tokyo 184--8501, Japan}\\[1mm]
{\em (4) Department of Physics, Tohoku University, Sendai 980--8578, Japan}

\ece

\vspace{6mm}

\begin{abstract}
  We calculate the ${\cal O}(\a_s)$ SUSY--QCD corrections to the widths 
  of stop and sbottom decays into Higgs bosons 
  within the Minimal Supersymmetric Standard Model.
  We give the complete analytical formulae paying particular attention 
  to the on--shell renormalization of the soft SUSY--breaking 
  parameters. 
  We also perform a detailed numerical analysis of both stop and 
  sbottom decays into all Higgs bosons $h^{0}$, $H^{0}$, $A^{0}$,
  and $H^{\pm}$. 
  We find that the SUSY--QCD corrections are significant, 
  mostly negative and of the order of a few ten percent. 
\end{abstract}

%------------------------------------------------------------------------
\section {Introduction}
%------------------------------------------------------------------------

Supersymmetry (SUSY) requires the existence of two scalar partners
$\sqL$ and $\sqR$ to each quark $q$. In the case of the
scalar partners
of the top quark one expects a large mixing between 
$\st_L$ and $\st_R$ due to the large top quark mass \cite{gh}.
The mixing of $\sb_L$ and $\sb_R$ may also be substantial if
$\tan\beta = v_{2}/v_{1}$ is large (where $v_{1}$ and $v_{2}$ are the
vacuum expectation values of the two Higgs doublets).
Strong mixing induces large mass differences between the lighter
mass eigenstate $\sq_1^{}$ and the heavier one $\sq_2^{}$, $\sq = \st$
or $\sb$. This implies in general a very complex decay pattern
of the heavier states. In addition to the ``conventional'' decays
into neutralinos, charginos, and gluinos
$(i,j = 1,2; \: k = 1, \ldots,4)$
\begin{equation}
  \st_{i} \to t\,\nt_k ,\,  b\,\chp_j ,  
  \qquad 
  \sb_{i} \to b\,\nt_k ,\,  t\,\chm_j ,
  \label{sqchdec}
\end{equation}
\begin{equation}
  \st_{i} \to t\,\sg ,  \qquad 
  \sb_{i} \to b\,\sg ,
  \label{sqsgdec}
\end{equation}
decays into vector bosons and Higgs particles can become kinematically
possible ($i,j=1,2$):
\begin{equation}
  \st_2 \to \st_1 Z^0 ,\;\; \st_i \to  \sb_j W^+\, ,
  \qquad 
  \sb_2 \to \sb_1 Z^0 ,\;\; \sb_i \to  \st_j W^-\, ,
  \label{sqWZdec}
\end{equation}
\begin{equation}
  \st_2 \to \st_1 (h^0, H^0, A^0) ,\;\; \st_i \to \sb_j H^+ ,
  \qquad 
  \sb_2 \to \sb_1 (h^0, H^0, A^0) ,\;\; \sb_i \to \st_j H^- .
  \label{sqhdec}
\end{equation}
All these squark decays were first discussed at tree--level in \cite{porod}
within the Minimal Supersymmetric Standard Model (MSSM) \cite{HaberKane}. 
A recent, more complete and systematic analysis of these decays at tree--level 
in \cite{squark2dec} revealed that the bosonic decays of 
Eqs.~(\ref{sqWZdec}) and (\ref{sqhdec}) can be dominant in a wide range
of the MSSM parameters due to the large Yukawa couplings and mixings
of $\st$ and $\sb$. This could have an important impact
on the search for $\st_{2}$ and $\sb_{2}$ and the determination of the 
MSSM parameters at future colliders. Therefore it is important
to study how SUSY--QCD corrections affect this tree--level result.\\
Within the last years SUSY--QCD corrections to a variety of processes
were calculated. 
For the decays of Eq.~(\ref{sqchdec}) this was done in 
\cite{sqchdecay,djouadihollik,beenakker1}, 
for the decays Eq.~(\ref{sqsgdec}) in \cite{beenakker1,beenakker2}, 
and for the decays Eq.~(\ref{sqWZdec}) in \cite{sqWZdecay}. 
The SUSY--QCD corrections for the decays into Higgs bosons, 
Eq.~(\ref{sqhdec}), were briefly discussed in \cite{karlsruhe}.  
The QCD corrections to the related Higgs boson
decays $(h^{0}, H^{0}, A^{0}) \to \sq_i^{} {\bar{\sq}}_{j}$ and
$H^{\pm} \to \sq_i^{} \bar{\sq}_{j}'$ were calculated 
in \cite{higgsdec,karlsruhe}.
A thorough study of the corrections to the decays Eq.~(\ref{sqhdec}), 
including a detailed numerical analysis, is, however, still missing 
in the literature. 

In this article we discuss the ${\cal O}(\alpha_{s})$
SUSY--QCD corrections to the decay widths of Eq.~(\ref{sqhdec})
in the on--shell renormalization scheme within the MSSM.
We give the complete formulas for these corrections and point out 
some subtleties which occur in the on--shell renormalization scheme.
Whereas a numerical analysis was made only for 
$\st_2\to\st_1+(h^0,A^0)$ in \cite{karlsruhe}, we perform a detailed analysis on stop and 
sbottom decays into all Higgs bosons $h^0$, $H^0$, $A^0$, and $H^\pm$.

%------------------------------------------------------------------------
\section {Tree--level formulae and notation}
%------------------------------------------------------------------------

We first summarize the tree--level results and our notation. 
The squark mass matrix in the basis $(\sqL, \sqR)$ 
is given by \cite{gh}
\beq
  {\cal M}_{\sq}^2 = 
  \left( \begin{array}{cc}  m^2_{\sq_L} & a_q m_q \\
                            a_q m_q     & m^2_{\sq_R}
  \end{array}\right) \;=\;
  (\R^{\sq})\kr
    \left( \begin{array}{cc}  m^2_{\sq_1} & 0 \\
                              0 & m^2_{\sq_2}
    \end{array}\right) \R^{\sq}
  \label{eq:massmat}
\eeq
with
\beqaa
  m^2_{\sq_L} &=& M^2_{\ti Q} 
    + m_Z^2 \cos 2\beta\,(I_{3L}^q - e_q\sin^2\t_W) + m_q^2, \\
  m^2_{\sq_R} &=& M^2_{\{\ti U,\ti D\}} 
    + m_Z^2 \cos 2\b\, e_q\sin^2\t_W + m_q^2, \\
  a_q &=& A_q - \mu\, \{ \cot\b,\,\tan\b \}
\eeqaa
for $\{$up, down$\}$ type squarks.  
$M_{\ti Q, \ti U, \ti D}$ and $A_{t,b}$ are soft SUSY--breaking 
parameters and $\mu$ is the Higgs mixing parameter in the superpotential.
$I_{3}^q$ and $e_q$ are the third component of the weak isospin and 
the electric charge of the quark $q$.
The squark mixing matrix ${\cal R}^{\sq}$ is 
\beq 
  {\cal R}^{\sq} = 
  \left(\baa{rr} \cth_{\sq} & \sth_{\sq} \\ 
                -\sth_{\sq} & \cth_{\sq} \eaa\right) 
  \label{eq:Rsq}
\eeq 
with $0\leq\tsq < \pi$ by convention. 
The weak eigenstates $\sqL$ and $\sqR$ are related to their 
mass eigenstates $\sqe$ and $\sqz$ (with $\msq{1}\leq\msq{2}$) by 
\beq
  {\sqe \choose \sqz} = {\cal R}^{\sq}\,{\sqL \choose \sqR}.
\eeq

In the ($\sq_1^{},\:\sq_2^{}$) basis the squark interaction with  
Higgs bosons $\Hk = \{h^0\!,\,H^0\!,\,A^0\!,\,H^\pm \}$ can be written 
as ($i,j=1,2$; $k=1\ldots 4$; $\a$ and $\b$ flavor indices) 
\beq
  {\cal L}_{\sq\sq H} = 
      G_{ijk}^{\,\a}\, H_{k}^{\dagger}\, \sq_j^{\b\dagger}\, \sq_i^\a .
\eeq
The couplings $G_{ijk}^{\alpha}$ are  
\beq  
  G_{ijk}^{\,\sq} = \left[\, 
  \R^{\sq}\,\hat{G}_k^{\,\sq}\, (\R^{\sq})^{\rm T} \,\right]_{ij}                 
  \hspace{6mm} (k=1,2,3),
  \label{eq:sqh0cop} 
\eeq
\beq
  G_{ij4}^{\,\st} = \left[\, 
  \R^{\st}\,\hat G_4^{}\, (\R^{\sb})^{\rm T}\,\right]_{ij}, 
  \hspace{6mm}
  G_{ij4}^{\,\sb} = \left[\, 
  \R^{\sb}\,(\hat G_4^{})^{\rm T}\,(\R^{\st})^{\rm T}\,\right]_{ij},
  \label{eq:sqH4cop}                 
\eeq
with $\hat G_k^{}$, $k=1\ldots 4$, being the respective couplings 
in the ($\sqL,\:\sqR$) basis as given in the Appendix.
Note that $G_{ij4}^{\,\sb} = G_{ji4}^{\,\st}$ and 
$G_{ij3}^{\,\sq} = \left[ \hat G_3^{\,\sq}\right]_{ij}$.  
\noi
The tree--level width of the decay $\sq^\a_i \to \sq^\b_j \Hk$,  
Fig.~1a, is thus given by \cite{porod}
\beq
  \Gamma^{0} (\sq^\a_i \to \sq^\b_j \Hk) =
  \frac{|G_{ijk}^{\,\a}|^2\: \kappa (m_i^2,m_j^2,m_H^2)}
       {16\pi\,m_i^3 }
\label{eq:Gtree}
\eeq
where $m_i \equiv m_{\sq^\a_i}$, $m_j \equiv m_{\sq^\b_j}$, 
$\mH \equiv m_{\!H_{\!k}}^{}$, and 
$\kappa (x,y,z) = [(x-y-z)^2 - 4yz]^{1/2}$. 
For $k=1,2,3$ we have of course $\a=\b$ and $i=2$, $j=1$. 
For $k=4$ we have $(\sq_{i}^{\alpha}, \sq_{j}^{\beta})$ =
$(\st_{i}, \sb_{j})$ or $(\sb_{i}, \st_{j})$.
In the following, we will omit flavor indices when possible 
(flavor $= \a$ if not given otherwise).

%------------------------------------------------------------------------
\section {SUSY--QCD corrections}
%------------------------------------------------------------------------

The ${\cal O}(\a_{s})$ corrected decay amplitude in the on--shell
renormalization scheme can be expressed as 
\beq
  G_{ijk}^{\mbox{\footnotesize corr}} = 
  G_{ijk} + \d G_{ijk}^{(v)} + \d G_{ijk}^{(w)} + \d G_{ijk}^{(c)}
\label{Gijkcorr}
\eeq
where the superscript $v$ denotes the vertex correction (Fig.~1b), 
$w$ the squark wave--function correction (Fig.~1c), 
and $c$ the counterterm due to 
the shift from the bare to the on--shell couplings. 
The ${\cal O}(\a_{s})$ corrected decay width $\G$ is then given by 
\beq
  \G = \G^{0} + \d\G^{(v)} + \d\G^{(w)} + \d\G^{(c)} + \d\G_{real}
  \label{eq:gencorr}
\eeq
with 
\beq
  \d\G^{(a)} =
  \frac{ \kappa (m_i^2,m_j^2,\mH^2)}{8\pi\,m_i^3 }\;\,
  \mbox{Re}\left\{ G_{ijk}^{\,*}\:\d G_{ijk}^{(a)}\right\} 
  \hspace{6mm}
  (a = v,\,w,\,c) 
\eeq
and $\d\G_{real}$ the correction due to real gluon emission 
(Fig.~1e)
which has to be included in order to cancel the infrared divergences.
We use dimensional reduction \cite{dimred} as regularization scheme. 
Analogous calculations were performed for the crossed channels of 
Higgs decays into squarks in \cite{higgsdec,karlsruhe}.

%--------------------------------
\subsection{Vertex corrections}
%--------------------------------

The vertex correction due to the gluon--squark--squark loop in Fig.~1b is
\beq
  \d G_{ijk}^{(v,g)} = \frac{\a_s}{3\pi}\: G_{ijk} \left[\,
    B_0(m_i^2,0,m_i^2) + B_0(m_j^2,0,m_j^2) 
    - B_0(m_{H_{k}}^2,m_i^2,m_j^2) + 2\,X\,C_0
  \,\right]
\eeq
with $X = m_i^2 + m_j^2 - m_{H_{k}}^2$. 
$B_0$ and $C_0$ are the standard two-- and three--point 
functions \cite{pave} for which we follow the conventions of \cite{denner}. 
In this case, $C_0 = C_0(m_i^2, m_{H_{k}}^2, m_j^2; \l^{2}, m_i^2, m_j^2)$. 
As usual, we introduce a gluon mass $\l$ to regularize the 
infrared divergence. 

The graph with the gluino--quark--quark loop in Fig.~1b
leads to
\beq
  \d G_{21\ell}^{(v,\sg)} 
  = -\frac{2}{3}\frac{\a_s}{\pi}\;\msg\,\cos 2\tsq\:s_\ell^{\a}\, 
    \Big[\,B_0(\msq{2}^2,\msg^2,m_q^2) + B_0(\msq{1}^2,\msg^2,m_q^2) + 
    (4\,m_q^2 - m_{\!H_\ell}^2)\,C_0 \,\Big] 
\label{eq:loopsgA}
\eeq
for the decays into $h^0$ and $H^0$ ($\ell = 1,2$),
\beqaa
  \d G_{213}^{(v,\sg)} &=& 
  -\frac{2}{3}\frac{\a_s}{\pi}\;s_3^{\a} \,\left\{
    \,m_q\sin 2\tsq 
    \left[\, B_0(\msq{2}^2,\msg^2,m_q^2) - B_0(\msq{1}^2,\msg^2,m_q^2)  
             + (\msq{2}^2 - \msq{1}^2)\,C_0
  \,\right] \right.\nn\\
  & & \hspace{22mm}\left. +\,\msg \left[\, 
    B_0(\msq{2}^2,\msg^2,m_q^2) + B_0(\msq{1}^2,\msg^2,m_q^2) 
    - m_{\!A}^2\,C_0 \,\right] \,\right\},
\label{eq:loopsgB}
\eeqaa
for the decay into $A^0$, and 
\beqaa
\d G_{ij4}^{(v,\sg)} &=& \frac{2}{3}\frac{\a_s}{\pi}\,\left\{\, 
  \left[\, (m_{q^\a} A_{11} + m_{q^\b} A_{22})\,s_4^\a
          +(m_{q^\a} A_{22} + m_{q^\b} A_{11})\,s_4^\b
  \,\right] B_0(m_{\!H^+}^2,m_t^2,m_b^2) \right. \nn\\
  & & \hspace{11mm} 
  +\,\left[\, (m_{q^\a} A_{11} - \msg A_{21})\,s_4^\a
             +(m_{q^\a} A_{22} - \msg A_{12})\,s_4^\b
  \,\right] B_0(m_i^2,\msg^2,m_{q^\a}^2) \nn\\ 
  & & \hspace{11mm} 
  +\,\left[\, (m_{q^\b} A_{22} - \msg A_{21})\,s_4^\a
             +(m_{q^\b} A_{11} - \msg A_{12})\,s_4^\b
  \,\right] B_0(m_j^2,\msg^2,m_{q^\b}^2) \nn\\ 
  & & \hspace{11mm} 
  +\,\left[\,m_{q^\b}\,(m_{q^\a}^2 - m_i^2 + \msg^2)\:
                       (A_{22}\,s_4^\a + A_{11}\,s_4^\b) \right. \nn\\
  & & \hspace{22mm} 
          +\,m_{q^\a}\,(m_{q^\b}^2 - m_j^2 + \msg^2)\:
                      (A_{11}\,s_4^\a + A_{22}\,s_4^\b)  \nn\\
  & & \hspace{22mm}   
          +\,\msg\,(m_{\!H^+}^2 - m_{q^\a}^2 - m_{q^\b}^2)\:
                  (A_{21}\,s_4^\a + A_{12}\,s_4^\b) \nn\\
  & & \hspace{22mm} \left. \left.  
          -\,2\msg\,m_{q^\a}\,m_{q^\b}\:(A_{12}\,s_4^\a + A_{21}\,s_4^\b)
  \right] C_0 \,\right\} 
\label{eq:loopsgC}  
\eeqaa
with $A_{nm} = \R_{in}^\a \R_{jm}^\b$ for the decay into a charged 
Higgs boson. \\
In Eqs.~(\ref{eq:loopsgA}) to (\ref{eq:loopsgC}) 
$C_{0} = C_{0}(m_i^2, m_{H_{k}}^2, m_j^2; \msg^{2}, m_{q^\a}^2, m_{q^\b}^2)$;  
$s_k^{q}$ are the Higgs couplings to quarks: 
\beq
  {\cal L}_{qqH} =
  s_1^q\,h^0\bar{q}\,q + s_2^q\,H^0\bar{q}\,q 
                       + s_3^q\,A^0\bar{q}\,\g^5\,q  
  +\,H^+\,\bar{t}\,(s_4^b\,P_R + s_4^t\,P_L)\,b 
  + H^-\,\bar{b}\, (s_4^t\,P_R + s_4^b\,P_L)\,t
\eeq
with
\beq 
  \begin{array}{llll} 
    s_1^t = -\,\frac{g\,m_t}{2\,m_W\sin\b}\,\cos\a, &
    s_2^t = -\,\frac{g\,m_t}{2\,m_W\sin\b}\,\sin\a, &
    s_3^t = i\,\frac{g\,m_t}{2\,m_W}\cot\b, &
    s_4^t = \,\frac{g\,m_t}{\rzw\,m_W}\cot\b, 
    \\[2mm]
    s_1^b = \frac{g\,m_b}{2\,m_W\cos\b}\,\sin\a, &
    s_2^b = -\,\frac{g\,m_b}{2\,m_W\cos\b}\,\cos\a, &
    s_3^b = i\,\frac{g\,m_b}{2\,m_W}\tan\b, &
    s_4^b = \,\frac{g\,m_b}{\rzw\,m_W}\tan\b. 
  \end{array}
\eeq

The vertex correction due to the four--squark interaction 
in Fig.~1b is
\beq
  \d G_{ijk}^{(v,\sq)} = -\frac{\a_s}{3\pi}
  \sum_{n,m=1,2}{\cal S}_{in}^{\a}\,{\cal S}_{jm}^{\b}\,G_{nmk}^{ }\,
  B_0(m_{H_{k}}^2,m_{\sq_m^\b}^2,m_{\sq_n^\a}^2)
\eeq
with 
\beq
  {\cal S}^\a_{in} =  \left( \begin{array}{rr}  
      \cos 2\tsq  & -\sin 2\tsq  \\
      -\sin 2\tsq & -\cos 2\tsq
  \end{array} \right)_{in}^\a
\eeq

%--------------------------------------
\subsection{Wave--function correction}
%--------------------------------------

The wave--function correction is 
\beq 
  \d G_{ijk}^{(w)} = 
  \onehf \left[ 
    \d\ti Z_{ii}^{\alpha} + \d\ti Z_{jj}^{\beta} \right] G_{ijk}^{\alpha}
    + \d\ti Z_{i'i}^{\alpha}\,G_{i'jk}^{\alpha} 
    + \d\ti Z_{j'j}^{\beta}\,G_{ij'k}^{\alpha},  
  \hspace{8mm} {\footnotesize
  \begin{array}{ll} i \not= i' \\ j \not= j' \end{array} } 
\eeq
where the $\ti Z_{nm}^{\alpha}$ are the squark wave--function renormalization 
constants for $\sq^{\alpha}$. 
They stem from the gluon, gluino, and squark loops
of Fig.~1c \footnote{The gluon loop due to the $\sq\sq gg$ interaction 
gives no contribution because it is proportional to 
$\l^2{\rm ln}\l \to 0$.} 
and are given by:
\beq
  \delta \ti Z_{nn}^{(g,\sg)} = 
    - \mbox{Re}\left\{\dot\Sigma_{nn}^{(g,\sg)}(\msq{n}^{2}) \right\}\,, 
  \quad
  \delta \ti Z_{n'n}^{(\sg,\sq)} =
  \frac{\mbox{Re}\left\{\Sigma_{n'n}^{(\sg,\sq)}(\msq{n}^{2})\right\} 
         }{\msq{n'}^2 - \msq{n}^2} \,, 
  \hspace{6mm} n \neq n' 
\eeq
with $\dot\Sigma_{nn} (m^2) = 
\partial\Sigma_{nn} (p^2)/\partial p^2 |_{p^2=m^2}$. 
The contribution due to gluon exchange is
\beq
  \dot\Sigma_{nn}^{(g)} (\msq{n}^2) = - \frac{2}{3}\frac{\alpha_s}{\pi} 
  \left[
     B_{0}(\msq{n}^{2}, 0, \msq{n}^{2}) 
     + 2\msq{n}^{2} \dot B_{0} (\msq{n}^{2}, \l^{2}, \msq{n}^{2})
  \right] ,
\eeq
and those due to gluino exchange are
\beqaa  
  \dot\Sigma_{nn}^{(\sg)} (\msq{n}^2) &=& \frac{2}{3}\frac{\a_{s}}{\pi} 
  \left[ B_0(\msq{n}^2, \msg^2,m_q^2) + (\msq{n}^2-m_q^2-\msg^2)\, 
         \dot B_0(\msq{n}^2, \msg^2, m_q^2) \right.\nn\\
  & & \left.\hspace*{12mm} 
    -\, 2 m_q\msg (-1)^n\sin 2\tsq\, \dot B_0 (\msq{n}^2, \msg^2,m_q^2) 
  \right] ,
\eeqaa
\beq
  \Sigma_{12}^{(\sg)}(\msq{n}^2) = \Sigma_{21}^{(\tilde g)}(\msq{n}^2) =
  \frac{4}{3}\frac{\a_s}{\pi}\:\msg m_q \cos 2 \tsq \,
  B_0(\msq{n}^2, \msg^2,m_q^2).
\eeq
The four--squark interaction gives
\beq
\Sigma_{12}\hsq (\msq{n}^2) = \Sigma_{21}\hsq (\msq{n}^2) = 
  \frac{\alpha_s}{6\pi}\sin 4\tsq
  \left[ A_{0}(m_{{\sq}_2}^2) - A_{0}(m_{{\sq}_1}^2) \,\right]
\eeq
where $A_{0}(m^{2})$ is the standard one--point function 
in the convention of \cite{denner}. 

%----------------------------------------------------
\subsection{Shift from bare to on--shell parameters}
%----------------------------------------------------

We next fix the shift from the bare to the on--shell couplings
$\d G_{ijk}^{(c)}$ in Eq.~(\ref{Gijkcorr}). 
We follow the procedure which was first given in \cite{higgsdec}.
From Eqs.~\eq{sqh0cop} and \eq{Rsq} we get for the squark decays 
into $h^0$ or $H^0$ ($\ell=1,2$)
\beqaa
  \d G_{21\ell}^{\,\sq (c)} &=& \left[\,
    \R^{\sq}\, \d \hat G_\ell^{\,\sq}\, (\R^{\sq})^{\rm T} 
    +\d\R^{\sq}\, \hat G_\ell^{\,\sq}\, (\R^{\sq})^{\rm T} 
    +\R^{\sq}\,   \hat G_\ell^{\,\sq}\, \d (\R^{\sq})^{\rm T} 
    \,\right]_{21} \nn \\
  &=& 
  \cos 2\tsq \,\left[\,\d \hat G_\ell^{\,\sq}\,\right]_{21}^{} 
  -( G_{11\ell}^{\,\sq} - G_{22\ell}^{\,\sq} )\, \d\tsq \,
\eeqaa
with $\d \hat G_\ell^{\,\sq}$ obtained by varying 
Eqs.~(\ref{eq:GLR1}--\ref{eq:GLR2}), e. g.
\beqaa
  \d \hat G_2^{\,\st} &=& -\smaf{g}{2 m_W^{} {\rm s}_\b} 
     \left( \begin{array}{cc}
        4 m_t\,{\rm s}_\a\,\d m_t 
             & \d (m_t A_t)\,{\rm s}_\a - \mu\,{\rm c}_\a\,\d m_t  \\
        \d (m_t A_t)\,{\rm s}_\a  - \mu\,{\rm c}_\a\,\d m_t 
             & 4 m_t\,{\rm s}_\a\,\d m_t
     \end{array} \right)\, , 
  \label{eq:dGLR2st}
  \\[3mm]
  \d \hat G_2^{\,\sb} &=& -\smaf{g}{2 m_W^{} {\rm c}_\b} 
     \left( \begin{array}{cc}
        4 m_b {\rm c}_\a\d m_b  
             & \d (m_b A_b) {\rm c}_\a - \mu {\rm s}_\a \d m_b \\
        \d (m_b A_b) {\rm c}_\a - \mu {\rm s}_\a \d m_b  
             & 4 m_b {\rm c}_\a \d m_b
     \end{array} \right)\, 
  \label{eq:dGLR2sb}
\eeqaa
with ${\rm s}_\b = \sin\b$, ${\rm c}_\b = \cos\b$,
${\rm s}_\a = \sin\a$, ${\rm c}_\a = \cos\a$, 
and $\a$ the Higgs mixing angle. 
$\d \hat G_{1}^{\,\sq}$ is obtained from
Eqs.~\eq{dGLR2st} and \eq{dGLR2sb} by: 
\beq
  \d \hat G_{1}^{\,\sq} = \left(
  \d \hat G_{2}^{\,\sq} \;\mbox{with}\; \a \to \a + \pi/2 \right)\, . 
\eeq
 
For the couplings to the $A^0$ boson we have explicitly 
\beq 
  \d G_{213}^{\,\sq\,(c)} = \smaf{ig}{2 m_W} \,\left[\,
    \d(m_q A_q)\, \{\cot\b, \tan\b\} + \mu\, \d m_q \,\right],
\eeq
where $\cot\b$ ($\tan\b$) is for $\sq = \st$ ($\sb$).

\noi 
For the decay $\st_i \to \sb_j\, H^+$ ($k = 4$) we get
\beq 
  \d G_{ij4}^{\,\st\,(c)} = 
  \left[\,\R^{\st}\, \d \hat G_4\, (\R^{\sb})^{\rm T} \,\right]_{ij}
  - (-1)^i\, G_{i'j4}^{\,\st}\, \d\t_{\st} 
  - (-1)^j\, G_{ij'4}^{\,\st}\, \d\t_{\sb} \, ,
  \hspace{8mm} {\footnotesize
  \begin{array}{ll} i \not= i' \\ j \not= j' \end{array} } 
\eeq
with 
\beq
  \d \hat G_4 = \smaf{g}{\sqrt{2}\,m_W^{}} 
  \left( \begin{array}{cc}
     2 m_b \d m_b\tan\b + 2 m_t \d m_t \cot\b  
     & \d (m_b A_b)\tan\b + \mu\, \d m_b \\
     \d (m_t A_t)\cot\b + \mu\,\d m_t   
     & 2(\d m_t m_b + m_t \d m_b)/\sin 2\b
  \end{array}\right) ,
\eeq
and analogously the expression
for $\sb_i \to \st_j\, H^-$ according to \eq{sqH4cop}.

$\d m_{q}$ is the shift from the bare to the pole mass of the quark 
$q$ and has two contributions (Fig.~1d). 
The gluon exchange contribution is
\beq
  \d m_q^{(g)} = - \frac{2}{3}\frac{\a_s}{\pi}\; 
    m_q \left[\, B_0(m_q^2, 0, m_q^2) - B_1(m_q^2, 0, m_q^2) - r/2\, 
        \right] ,
\label{eq:dmqg}
\eeq
and the gluino contribution is
\beqaa
  \d m_q^{(\ti g)} &=& - \frac{\a_s}{3\pi}\, \left\{\, 
  m_q\,\Big[\, B_1(m_q^2, \msg^2, \msq{1}^2) 
         + B_1(m_q^2, \msg^2, \msq{2}^2) \,\Big] \right. \nn \\
  & & \hspace{13mm} \left.
  +\,\msg \sin 2\tsq\,\Big[\,B_0(m_q^2, \msg^2, \msq{1}^2) 
                - B_0(m_q^2, \msg^2, \msq{2}^2) \,\Big]\, \right\}\, .
\eeqaa
The parameter $r$ in Eq.~\eq{dmqg} exhibits the dependence on the 
regularization scheme. As $r$ does not cancel in the final result, we 
have to use dimensional reduction \cite{dimred} ($r=0$) which preserves 
supersymmetry at least at two--loop order. 

For the renormalization of $\,m_q A_q$ we define the on--shell 
parameters $M_{\ti Q, \ti U, \ti D}$ and $A_{t,b}$ in terms of squark 
pole masses $\msq{i}$ and on--shell mixing angels $\tsq$ (which will 
be defined below) using the tree--level relations \eq{massmat} 
to \eq{Rsq}. We thus get \cite{higgsdec}
\beq 
  \d (m_q A_q) = 
  \onehf\,(\d \msq{1}^2 - \d \msq{2}^2) \sin 2\tsq 
  + (\msq{1}^2 - \msq{2}^2) \cos 2\tsq\,\d\tsq
  + \mu\, \{\cot\b,\tan \b\}\,\d m_q .
\label{eq:dmqaq}
\eeq
where again $\cot\b\;(\tan\b)$ is for $\sq = \st\;(\sb)$. 
$\d\msq{i}^2$ is given by
\beq
  \d\msq{i}^2 = {\rm Re} \left[\,
  \Sigma_{ii}^{(g)} (\msq{i}^2) + \Sigma_{ii}^{(\sg)} (\msq{i}^2) + 
  \Sigma_{ii}^{(\sq)} (\msq{i}^2)
  \,\right]
\eeq
with
\beqaa
  \Sigma_{ii}^{(g)}(\msq{i}^2) &=&
    -\frac{2}{3}\frac{\a_s}{\pi}\,\msq{i}^2\, 
    \{ 2B_0(\msq{i}^2, 0,\msq{i}^2) + B_1(\msq{i}^2, 0,\msq{i}^2) 
    \}, \\ 
  \Sigma_{ii}^{(\sg)}(\msq{i}^2) 
  &=& -\frac{4}{3}\frac{\a_s}{\pi} \:\{ 
         A_0(m_q^2) + \msq{i}^2 B_1(\msq{i}^2,\msg^2,m_q^2) \nn\\
  & & \hspace{16mm} +\; [ \,
      \msg^2 + (-)^i\,\msg m_q \sin 2\tsq\,]\,B_0(\msq{i}^2,\msg^2,m_q^2) 
    \} , \\
  \Sigma_{ii}^{(\sq)}(\msq{i}^2) &=& \frac{\a_s}{3\pi} \:
    \{ \cos^2 2\tsq \, A_0(\msq{i}^2) + \sin^2 2\tsq \, A_0(\msq{i'}^2) \} ,
    \hspace{4mm} i \not= i'.
\eeqaa
We also have to renormalize the squark mixing angle. 
This problem was first solved in \cite{helmut}:  
there the counterterm $\d\tsq$ was fixed in the process 
$e^+e^-\to \sq_1^{}\bar{\sq}_2^{}$ such that it cancels the 
off--diagonal part of the squark wave--function corrections. 
$\d\tsq = \d\tsq\hsq + \d\tsq\hsg$ is thus given by  
\beq  
  \d\tsq\hsq = \frac{\a_s}{6\pi} \,
  \frac{\sin 4\tsq}{\msq{1}^2 - \msq{2}^2}
  \left[ A_{0}(\msq{2}^2) - A_{0}(\msq{1}^2) \,\right] ,
\eeq
\beq  
  \d\tsq\hsg = \frac{\a_s}{3\pi}\,
  \frac{m_{\sg} m_q}{I_{3L}^q (m_{\sq_1}^2 - m_{\sq_2}^2)}
  \left[ B_{0}(m_{\sq_2}^2,m_{\sg}^2,m_q^2)\,\ti v_{11} -
         B_{0}(m_{\sq_1}^2,m_{\sg}^2,m_q^2)\,\ti v_{22}\,\right] , 
\eeq
where $\ti v_{ij}$ comes from the $Z^{0}\sq_{i}^{}\sq_{j}^{*}$ 
couplings,
$\ti v_{11}= 4(I^q_{3L} \cos^2 \tsq - s_W^2 e_q)$ and 
$\ti v_{22}= 4(I^q_{3L} \sin^2 \tsq - s_W^2 e_q)$ with 
$s_W^2 = \sin^2\t_W^{}$. 
We will use this scheme in what follows. 
There are also other possibilities of defining the on--shell squark 
mixing angle. 
In \cite{karlsruhe}, for instance, $\d\tsq$ was fixed such that 
%the renormalized self--energy of the lighter mass eigenstate remains diagonal. 
the renormalized self-energy of the squarks remains diagonal
on the $\sq_1$ mass shell.
Similar conditions were used in \cite{beenakker3,sola}. 
For comparison we list the counterterms $\d\tsq$ in these schemes
in terms of the squark self--energy $\Sigma_{12}$: 
\beqaa
  \d\tsq([15]) &=&
  \frac{1}{\ti v_{11}-\ti v_{22}}
  \frac{\mbox{Re}\left\{
   \ti v_{11}\Sigma_{12}(\msq{2}^{2})
    -\ti v_{22}\Sigma_{12}(\msq{1}^{2})   \right\} }
    {\msq{1}^2 - \msq{2}^2} \,, \qquad {\rm (used~in~this~paper)}\nn\\
  \d\tsq([10]) &=&
  \frac{\mbox{Re}\left\{\Sigma_{12}(\msq{1}^{2})\right\}
         }{\msq{1}^2 - \msq{2}^2} \,, \nn\\
  \d\tsq(Q^2)([16]) &=&
  \frac{\mbox{Re}\left\{\Sigma_{12}(Q^2)\right\}
         }{\msq{1}^2 - \msq{2}^2} \,, \nn\\
  \d\tsq([17]) &=&
  \frac{1}{2}\frac{\mbox{Re}\left\{
     \Sigma_{12}(\msq{1}^{2})+\Sigma_{12}(\msq{2}^{2}) \right\}
         }{\msq{1}^2 - \msq{2}^2} \,.
\eeqaa
The differences between them are ultraviolet finite.
In Fig.~\ref {fig:deltatheta} we compare $\delta\t_{\st}$ of 
\cite{karlsruhe,beenakker3,sola} to that of our scheme (\cite{helmut}).
As can be seen, the numerical differences between the various 
schemes are very small ($<1\%$). 
%Numerically, the differences between the 
%various schemes are very small, typically of ${\cal O}(10^{-2})$. 
 
There is yet another subtlety that has to be taken into account: 
At tree--level and in the $\overline{{\rm DR}}$ renormalization scheme 
SU(2)$_L$ symmetry requires that the parameter $M_{\ti Q}$ in 
the $\st$ and $\sb$ mass matrices have the same value. 
This is, however, not the case at loop--level in the on--shell scheme 
due to different shifts $\d M_{\ti Q}^2$ in the $\st$ and in the 
$\sb$ sectors \cite{higgsdec,yamada}. 
In this paper we take 
$M_{\ti Q}^2(\sq) = \msq{1}^2 \cos^2\t_{\sq} + 
\msq{2}^2 \sin^2\t_{\sq} 
- m_Z^2 \cos 2\b\, (I_{3L}^{q} - e_{q} \sW) - m_q^2$ 
as the on--shell parameter in the $\sq$ sector
$(\sq = \st, \sb)$. 
This leads to a shift of $M_{\ti Q}^2$
\cite{higgsdec}:
\beq
  M_{\ti Q}^2(\sb) = 
  M_{\ti Q}^2(\st) + \d M_{\ti Q}^2(\st) - \d M_{\ti Q}^2(\sb) 
\label{eq:mQshift}
\eeq
with
\beq
  \d M_{\ti Q}^2(\sq) = 
  \d\msq{1}^2 \cos^2\tsq + \d\msq{2}^2 \sin^2\tsq - 
  (\msq{1}^2-\msq{2}^2) \sin 2\tsq\,\d\tsq - 2 m_q \d m_q .
\label{eq:dmQshift}
\eeq
The underlying SU(2)$_L$ symmetry is reflected in the fact that the 
shift $\d M_{\ti Q}^2(\st) - \d M_{\ti Q}^2(\sb)$ is finite. 

%--------------------------------
\subsection{Real gluon emission}
%--------------------------------

In order to cancel the infrared divergence we 
include the emission of real (hard and soft) gluons (Fig.~1e): 
\beq
  \d\G_{real} = \G(\sq_i^\a \to \sq_j^\b\,\Hk\, g) 
  = -\frac{\a_s\,|G_{ijk}^{\,\a}|^2}{3\pi^2 m_i}\: \left[\, 
    I_0 + I_1 + m_i^2\,I_{00} + m_j^2\,I_{11} + X\,I_{01} \right] \, .
\eeq
Again, $X = m_i^2 + m_j^2 - m_{H_{k}}^2$.
The phase space integrals $I_{n}$, and $I_{nm}$ 
have $(m_i, m_j, m_{H_{k}})$ as arguments and are given in \cite{denner}.

\bigskip

We have checked explicitly that the corrected decay width $\Gamma$ of  
Eq.~\eq{gencorr} is ultraviolet and infrared finite.

%------------------------------------------------------------------------
\section{Numerical results and discussion}
%------------------------------------------------------------------------

Let us now turn to the numerical analysis.  
As input parameters we use $\mst{1}$, $\mst{2}$, $\cos\t_{\st}$,  
$\tan\beta$, $\mu$, $m_A$, and $\msg$. 
From these we calculate the values of the 
soft SUSY--breaking parameters $M_{\ti Q}(\st)$, $M_{\ti Q}(\sb)$, 
$M_{\ti U, \ti D}$, and $A_{t,b}$ according to 
Eqs.~\eq{massmat} to \eq{Rsq} together with
Eqs.~\eq{mQshift} and \eq{dmQshift}, 
taking $M_{\ti D}=1.12\,M_{\ti Q}(\st)$ and $A_b=A_t$ 
for simplicity to fix the sbottom sector.   
Moreover, we take $m_t = 175$ GeV, $m_b = 5$ GeV, $m_Z^{} = 91.2$ GeV, 
$\sin^2\t_W = 0.23$, $\a(m_Z^{}) = 1/129$, and $\a_s(m_Z^{}) = 0.12$. 
For the running of $\a_{s}$ we use $\a_{s}(Q) = 
12\pi/[(33 - 2\,n_{f})\,\ln (Q^{2}/\Lambda^{2}_{n_{\!f}})]$ 
with $n_{\!f}$ the number of quark flavors. 
We take $\a_{s} = \a_{s}(m_{\sq^{\alpha}_{i}})$ for the
$\sq^{\alpha}_{i}$ decay except for the shift in 
Eqs.~\eq{mQshift} and \eq{dmQshift} for which we take 
$\a_{s} = \a_{s}(M_{\tilde Q}(\st))$.  
For the radiative corrections to the $h^0$ and $H^0$ masses and their 
mixing angle $\a$ ($-\frac{\pi}{2}\leq\a<\frac{\pi}{2}$ by convention) 
we use the formulae of \cite{mh0}\footnote{Notice that \cite{mh0,mhc} 
have the opposite sign convention for the parameter $\mu$.}; 
for those to $m_{H^+}$ we follow \cite{mhc}. 
In order to respect the experimental mass bounds from LEP2 
\cite{lep2} and Tevatron \cite{tevatron} we impose 
$m_{h^0} > 90\gev$, $m_{\st_1,\sb_1} > 85\gev$, and $\msg>300\gev$. 
Moreover, we require $\d\rho\,(\st-\sb) < 0.0012$ \cite{drhonum} 
from electroweak precision measurements 
using the one--loop formulae of \cite{Drees-Hagiwara} and 
$A_t^2 < 3\,(M_{\ti Q}^2(\st) + M_{\ti U}^2 + m_{H_2}^2)$, 
$A_b^2 < 3\,(M_{\ti Q}^2(\sb) + M_{\ti D}^2 + m_{H_1}^2)$ with 
$m_{H_2}^2=(m_{A}^2+m_{Z}^2)\cos^2\b-\frac{1}{2}\,m_Z^2$ 
and $m_{H_1}^2=
(m_{A}^2+m_{Z}^2)\sin^2\b-\frac{1}{2}\,m_Z^2$ \cite{Deren-Savoy} 
from tree--level vacuum stability. 

As a reference point we take 
$\mst{1}=250$ GeV, $\mst{2}=600$ GeV, 
$\cos\t_{\st}=0.26$ ($\t_{\st}\simeq 75^\circ$), 
$\tan\b=3$, $\mu=550$ GeV, $m_A=150$ GeV, and $\msg=600$ GeV. 
This leads to $\msb{1}=564$ GeV 
\footnote{Notice that at tree--level one has $\msb{1}=560$ GeV because 
  $M_{\ti Q}=558$ GeV for both the $\st$ and $\sb$ mass matrices; 
  at $\Oas$, however, one gets $M_{\ti Q}(\st)=558$ GeV and 
  $M_{\ti Q}(\sb)=563$ GeV.}, 
$\msb{2}=627$ GeV, $\cos\t_{\sb}=0.99$, $A_{t,b}=-243$ GeV, 
$m_{h^0}=100$ GeV, $m_{H^0}=162$ GeV, 
$\sin\a = -0.58$, and $m_{H^+}=164$ GeV. 
Thus $\st_2$ can decay into $\st_1 + (h^0,H^0,A^0)$, and 
$\sb_{1,2}$ can decay into $\st_1 H^-$.

We first discuss the parameter dependence of the widths of
$\st_2$ decays into neutral Higgs bosons by varying
one of the input parameters of the reference point. 
We define the SUSY--QCD corrections as the 
difference between the corrected width $\Gamma$ of 
Eq.~\eq{gencorr} and the tree--level width $\Gamma^{0}$ of
Eq.~\eq{Gtree}.\\
Figure \ref{fig:mst2} shows the tree--level and the SUSY--QCD 
corrected widths of the decays $\st_2\to\st_1+(h^0,H^0,A^0)$ 
as a function of $\mst{2}$. 
The relative corrections $\d\G/\G^0 \equiv (\G-\G^0)/\G^0$ are about 
$-10\%$ for the decay into $h^0$ and $-9\%$ to $-62\%$
for the decay into $A^0$. 
The corrections for $\st_2\to\st_1 H^0$ are 
$-9\%$, $-45\%$, and $+45\%$ for 
$\mst{2} =420, 670$, and $900$~GeV, respectively. 
The spikes in the corrected decay widths for $\mst{2}=775$ GeV are due 
to the $\st_2\to t\sg$ threshold.
The different shapes of the decay widths can be understood by the 
wide range of the parameters entering the Higgs couplings to stops. 
In the range $\mst{2}=300$ GeV to 900 GeV, we have 
$A_t=144$ GeV to $-889$ GeV and $\sin\a=-0.52$ to $-0.73$ 
($m_{h^0}=81$ GeV to 114 GeV, and $m_{H^0}=163$ GeV to 170 GeV). \\
Figure \ref{fig:costh} shows the $\cos\theta_{\st}$ 
dependence of the tree--level and the 
corrected widths of $\st_2\to\st_1+(h^0,H^0,A^0)$ decays. 
Again the shapes of the decay widths reflect
their dependence on the underlying SUSY parameters in a 
characteristic way. 
In particular we have 
$A_t= 1033$, $183$, $-666$~GeV and 
$\sin\a= -0.748$, $-0.565$, $-0.726$ for 
$\cos\t_{\st}=-0.7$, $0$, $0.7$, respectively. 
Apart from the points where the tree--level 
decay amplitudes vanish the relative corrections 
range from $-40\%$ to $20\%$. \\
In \fig{mA} we show the tree--level and the SUSY--QCD corrected decay 
widths as a function of $m_A$. 
For $m_A=100$, 200, 300 GeV we have 
$m_{h^0}=85$, 104, 105 GeV, 
$m_{H^0}=128$, 207, 304 GeV, and 
$\sin\a=-0.87$, $-0.45$, $-0.37$, respectively. 
The corrections to $\G^0(\st_2\to\st_1 h^0)$ range 
from $-15\%$ to $-7\%$ for $m_A=100$ GeV to 400 GeV. 
Those to $\G^0(\st_2\to\st_1 H^0)$ are $-50\%$ to $-22\%$ 
for $m_A\gsim 114$ GeV, and those to $\G^0(\st_2\to\st_1 A^0)$ are 
about $-25\%$. \\
As for the dependence on the gluino mass, as 
can be seen in \fig{msg}, the gluino decouples very slowly: 
in the range $\msg=300$ GeV to 1500 GeV $\d\G/\G^0$ varies from 
($-9\%$, $-37\%$, $-28\%$) to ($-7\%$, $-16\%$, $-14\%$) 
for the decays 
$\st_2\to\st_1+(h^0,H^0,A^0)$, apart from the 
$\st_2\to t\sg$ threshold at $\msg=425$ GeV. \\
As for the dependence on $\tan\b$, we get 
$\G(\st_2\to\st_1 h^0) = 2.68$, $2.09$, $1.42$~GeV with 
$\d\G/\G^0 \simeq -10\%$, $-7\%$, $-5\%$ for $\tan\b= 3, 10$, $30$,
respectively. 
Likewise, we get 
$\G = 0.67$, $1.61$, $2.45$~GeV with 
$\d\G/\G^0 \simeq -27\%$, $-19\%$, $-17\%$ 
for the decay into $H^0$ and 
$\G = 2.1$, $2.64$, $2.92$~GeV with 
$\d\G/\G^0 \simeq -22\%$, $-19\%$, $-18\%$
for the decay into $A^0$, respectively.

Let us now turn to the sbottom decays.  
We start again from the reference point given above. 
For the decay $\sb_1 \to \st_1 H^-$ we get 
$\G=3.88$ GeV with $\d\G/\G^0=-24\%$, and 
for the decay $\sb_2 \to \st_1 H^-$ we get $\G=0.08$ GeV with   
$\d\G/\G^0=+87\%$.
As in our examples the width of the latter decay 
is usually quite small (because $\sb_2\simeq\sb_R$ and
$\st_1 \sim \st_R$) 
we will discuss only the parameter dependence of the $\sb_1$ decay. \\
Figure~\ref{fig:sbmst1} shows the tree--level and the SUSY--QCD 
corrected widths of this decay as a function of $\mst{1}$. 
The SUSY--QCD corrections are about $-25\%$.
Notice that at tree--level we have
$\msb{1}=556\gev$ to 566 GeV for $\mst{1}=85$~GeV to 400~GeV,
whereas at $\Oas$ we have $\msb{1}=561\gev$ to 570 GeV. 
Therefore, the thresholds at tree--level and one--loop level are 
slightly different. \\
The dependence on the stop mixing angle is shown in \fig{sbcosth} 
for $\tan\b=3$ and 10, and the other parameters as given above.
(For $|\cos\t_{\st}|\gsim 0.72$ the decay $\sb_1\to\st_1 H^-$ is 
kinematically not allowed.)
In case of $\tan\b=3$, the SUSY--QCD corrections range  
from about $-40\%$ to $26\%$, 
with $\d\G/\G^0 > 0$ for $\cos\t_{\st} \lsim -0.6$. 
In case of $\tan\b=10$ $\d\G/\G^0$ is much larger. 
For $\cos\t_{\st} \gsim 0.5$ and $\tan\b=10$ 
we even get a negative corrected decay width. 
This is mainly due to a large contribution stemming from the term 
$\d(m_b A_b)\sim \mu\tan\b \,\d m_b$ of Eq.~\eq{dmqaq} 
and was already mentioned in \cite{higgsdec}. 
The same problem can occur for the decays $\sb_2\to\sb_1+(h^0,H^0,A^0)$ 
and $\st_2\to\sb_1 H^+$ which may be important for large $\tan\b$  
(due to the large bottom Yukawa coupling and the 
large $\sb_1$--$\sb_2$ mass splitting).\\ 
We have also studied the dependence on $m_A$. 
In the case $\tan\b=3$ (10) we have found that 
$\d\G/\G^0(\sb_1\to\st_1 H^-) \sim -20\%$ $(-40\%)$ 
for $m_A=100$ GeV to 285 GeV and the other parameters as given above.\\
As for the dependence on the gluino mass, 
$\d\G/\G^0(\sb_1\to\st_1H^-)$ ranges from $-26\%$ to $-14\%$ 
($-47\%$ to $-39\%$) for $\msg=300$ GeV to 1500 GeV and $\tan\b=3$ (10).

In \cite{karlsruhe} a numerical analysis for the decays 
$\st_2\to \st_1 + (h^0,A^0)$ was made. 
Whereas we agree with their figure for $\st_2\to\st_1 A^0$, 
we find a difference of about 10\% in both the tree--level and the 
corrected widths of $\st_2\to\st_1 h^0$. This may be due to a different 
treatment of the radiative corrections to the $h^0$ mass and
mixing angle $\alpha$. \footnote{We thank A. Djouadi and W. Hollik 
for correspondence on this point.}

%------------------------------------------------------------------------
\section{Conclusions}
%------------------------------------------------------------------------

We have calculated the ${\cal O}(\a_s)$ SUSY--QCD 
corrections to the decay widths of 
$\sq_2^{}\to\sq_1^{}+(h^0,H^0,A^0)$ and $\sq_i^{}\to\sq_j' H^\pm$~  
($\sq=\st,\,\sb$) in the on--shell scheme. 
We have taken into account appropriate shifts for the soft 
SUSY--breaking parameters defined in terms of on--shell 
squark masses and mixing angles. 
It has turned out that the SUSY--QCD corrections are 
mostly negative and of the order of a few ten percent 
and should therefore be taken into account. 
 
%------------------------------------------------------------------------
\section*{Acknowledgements}
%------------------------------------------------------------------------

The work of A.B., H.E., S.K., W.M., and W.P. was supported by 
the ``Fonds zur F\"orderung der wissenschaftlichen Forschung'' 
of Austria, project no. P10843--PHY.
The work of Y.Y. was supported in part by the Grant--in--aid for Scientific
Research from the Ministry of Education, Science, and Culture of
Japan, No.~10740106, and by Fuuju--kai Foundation.

%------------------------------------------------------------------------
\section*{Appendix: Squark -- Squark -- Higgs Couplings}
%------------------------------------------------------------------------

(a) squark -- squark -- $h^0$
\beq
\hat G_1^{\,\sq} = \left(\! \begin{array}{cc}
  \frac{g\,m_Z^{}}{{\rm c}_W^{}}\,C_L^{}\,{\rm s}_{\a+\b} 
                        -\rzw\;m_q\,h_q \poss{{\rm c}_\a}{-{\rm s}_\a} 
  & -\frac{1}{\rzw}\,h_q\,\big( A_q \poss{{\rm c}_\a}{-{\rm s}_\a} 
                                +\mu\poss{{\rm s}_\a}{-c_\a}\big) 
  \\[4mm]
  -\frac{1}{\rzw}\,h_q\,\big( A_q \poss{{\rm c}_\a}{-{\rm s}_\a} 
                              +\mu\poss{{\rm s}_\a}{-c_\a}\big) 
  & \frac{g\,m_Z^{}}{{\rm c}_W^{}}\,C_R^{}\,{\rm s}_{\a+\b} 
                         -\rzw\,m_q\,h_q \poss{{\rm c}_\a}{-{\rm s}_\a} 
\end{array}\! \right) 
\label{eq:GLR1}
\eeq

\noi 
for {\small $\Big\{\!\begin{array}{c} \mbox{\footnotesize up} \\[-1mm]
\mbox{\footnotesize down} \end{array} \!\Big\}$}
type squarks respectively. 
We use the abbreviations ${\rm c}_W^{} = \cos\tW$, 
${\rm s}_\a$ = $\sin\a$,\\[-2mm]
${\rm c}_\a$ = $\cos\a$, 
${\rm s}_{\a+\b} = \sin(\a\!+\!\b)$,  
$C_L^{} = I_{3L}^q\!-\!e_q\sin^2\t_W^{}$, and 
$C_R^{} = e_q\sin^2\t_W^{}$. $\alpha$ is the mixing angle in the CP even
neutral Higgs boson sector.
$h_q$ are the Yukawa couplings:   
\beq
  h_{t} = \frac{g\,m_t}{\sqrt{2}\:m_{W}\sin\b}, \hspace{8mm} 
  h_{b} = \frac{g\,m_b}{\sqrt{2}\:m_{W}\cos\b}.
\eeq

\bigskip

(b) squark -- squark -- $H^0$
\beq
\hat G_2^{\,\sq} = \left(\! \begin{array}{cc}
  -\frac{g\,m_Z^{}}{{\rm c}_W^{}}\,C_L^{}\,{\rm c}_{\a+\b} 
                        -\rzw\;m_q\,h_q \poss{{\rm s}_\a}{{\rm c}_\a} 
  & -\frac{1}{\rzw}\,h_q\,\big( A_q \poss{{\rm s}_\a}{{\rm c}_\a} 
                                -\mu\poss{{\rm c}_\a}{s_\a}\big) 
  \\[4mm]
  -\frac{1}{\rzw}\,h_q\,\big( A_q \poss{{\rm s}_\a}{{\rm c}_\a} 
                              -\mu\poss{{\rm c}_\a}{s_\a}\big) 
  & -\frac{g\,m_Z^{}}{{\rm c}_W^{}}\,C_R^{}\,{\rm c}_{\a+\b} 
                        -\rzw\,m_q\,h_q \poss{{\rm s}_\a}{{\rm c}_\a} 
\end{array}\! \right) 
\label{eq:GLR2}
\eeq

\bigskip

(c) squark -- squark -- $A^0$
\beq
\hat G_3^{\,\sq} = i\,\smaf{g\,m_q}{2\,m_W} 
  \left(\! \begin{array}{cc}
     0 & - A_q\poss{\cot\b}{\tan\b} - \mu  
    \\[3mm]
     A_q\poss{\cot\b}{\tan\b} + \mu & 0
  \end{array}\! \right) 
\label{eq:GLR3}
\eeq

\bigskip

(d) squark -- squark -- $H^\pm$
\beq
\hat G_4^{ } = \smaf{g}{\rzw\,m_W^{}}
  \left(\! \begin{array}{cc}
    m_b^2\tan\b + m_t^2\cot\b - m_W^2\sin 2\b
      & m_b\,(A_b\tan\b + \mu) 
    \\[3mm]
    m_t\,(A_t\cot\b + \mu)  
      & \frac{2 m_t m_b}{\sin 2\b}
\end{array}\! \right) 
\label{eq:GLR4}
\eeq

%------------------------------------------------------------------------
% references
%------------------------------------------------------------------------

%------------------------------------------------------------------------
% Figures
%------------------------------------------------------------------------

\setlength{\unitlength}{1mm}

%------------------------------------------------------------------------
% Figure 1: Page with Feynman graphs:

\begin{figure}[h]  
\bce
\begin{picture}(125,165)
\put(0.5,15){\mbox{\psfig{file=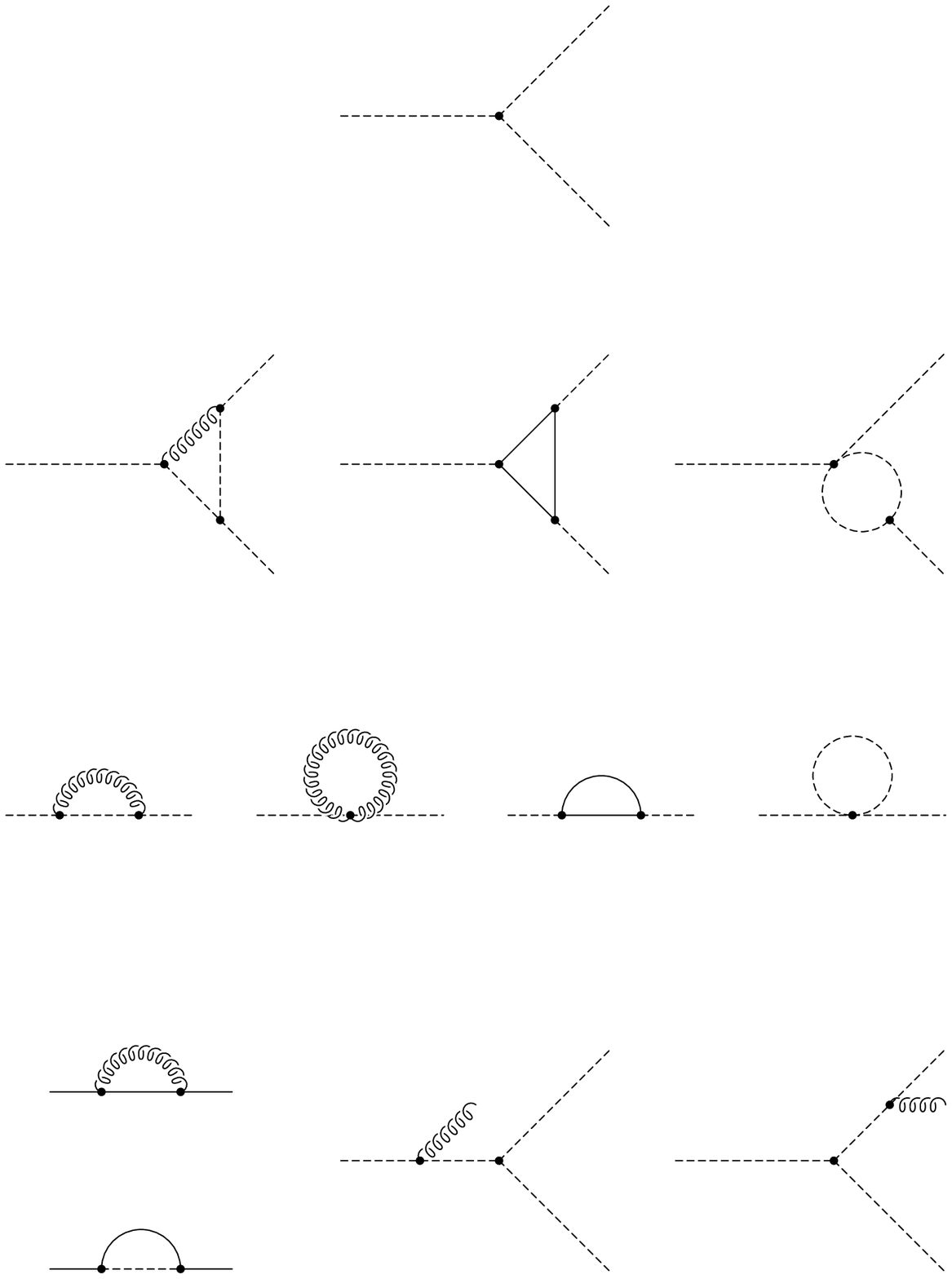,height=156mm}}}
%\put(0,0){\framebox(130,180){}}
% tree-level
\put(40,157){\makebox(0,0)[br]{$\sq_i^\a$}}
\put(77,170){\makebox(0,0)[bl]{$\sq_j^\b$}}
\put(77,142){\makebox(0,0)[bl]{$\Hk$}}
\put(60,139){\makebox(0,0)[tc]{\bf{(a)}}}
%
% gluon vertex corr.
\put(4,110){\makebox(0,0)[br]{$\sq_i^\a$}}
\put(35,127){\makebox(0,0)[bl]{$\sq_j^\b$}}
\put(34,99){\makebox(0,0)[bl]{$\Hk$}}
\put(22,120.5){\makebox(0,0)[br]{$g$}}
\put(23,107){\makebox(0,0)[br]{$\sq_i^\a$}}
\put(29,114){\makebox(0,0)[bl]{$\sq_j^\b$}}
% gluino vertex corr.
\put(45,110){\makebox(0,0)[br]{$\sq_i^\a$}}
\put(76,127){\makebox(0,0)[bl]{$\sq_j^\b$}}
\put(76,99){\makebox(0,0)[bl]{$\Hk$}}
\put(64,119.5){\makebox(0,0)[br]{$\sg$}}
\put(64,110){\makebox(0,0)[tr]{$q^\a$}}
\put(70,114){\makebox(0,0)[bl]{$q^\b$}}
% squark vertex corr.
\put(86,110){\makebox(0,0)[br]{$\sq_i^\a$}}
\put(117,127){\makebox(0,0)[bl]{$\sq_j^\b$}}
\put(117,99){\makebox(0,0)[bl]{$\Hk$}}
\put(99.5,109.5){\makebox(0,0)[tr]{$\sq_n^\a$}}
\put(111.5,113){\makebox(0,0)[bl]{$\sq_m^\b$}}
\put(60,95){\makebox(0,0)[ct]{\bf{(b)}}}
%
% squark wave-function correction
% gluon 1
\put(0,67.5){\mbox{$\sq_i^{}$}}
\put(20,67.5){\mbox{$\sq_i^{}$}}
\put(11,67.5){\mbox{$\sq_i^{}$}}
\put(11,79){\mbox{$g$}}
% gluon 2
\put(32,67.5){\mbox{$\sq_i^{}$}}
\put(50,67.5){\mbox{$\sq_i^{}$}}
\put(48,81){\mbox{$g$}}
% gluino 
\put(62,67.5){\mbox{$\sq_i^{}$}}
\put(82,67.5){\mbox{$\sq_j^{}$}}
\put(72,67.5){\mbox{$q$}}
\put(72.3,78.2){\mbox{$\sg$}}
% squark 
\put(93,67.5){\mbox{$\sq_i^{}$}}
\put(112.3,67.5){\mbox{$\sq_j^{}$}}
\put(110,80){\mbox{$\sq^{}$}}
\put(60,60){\makebox(0,0)[ct]{\bf{(c)}}}
%
% quark wave-function correction
% gluon 
\put(2.5,36.7){\mbox{$q$}}
\put(29.5,36.7){\mbox{$q$}}
\put(15.5,34){\mbox{$q$}}
\put(15.7,45.3){\mbox{$g$}}
% gluino 
\put(2.5,14.8){\mbox{$q$}}
\put(29.5,14.8){\mbox{$q$}}
\put(15.5,11.5){\mbox{$\sq^{}$}}
\put(16,22.5){\mbox{$\sg$}}
\put(17,6){\makebox(0,0)[ct]{\bf{(d)}}}
%
% real gluon emission
%
\put(40.3,28.5){\makebox(0,0)[br]{$\sq_i^\a$}}
\put(76.5,42.5){\makebox(0,0)[bl]{$\sq_j^\b$}}
\put(76,14){\makebox(0,0)[bl]{$\Hk$}}
\put(59,37){\makebox(0,0)[bl]{$g$}}
\put(81.3,28.5){\makebox(0,0)[br]{$\sq_i^\a$}}
\put(117,42.5){\makebox(0,0)[bl]{$\sq_j^\b$}}
\put(117,14){\makebox(0,0)[bl]{$\Hk$}}
\put(117,35.5){\makebox(0,0)[bl]{$g$}}
\put(79,6){\makebox(0,0)[ct]{\bf{(e)}}}
\end{picture}
\ece
\caption{Feynman diagrams relevant for the ${\cal O}(\a_{s})$ SUSY-QCD 
  corrections to squark decays into Higgs bosons. }
\end{figure}

%------------------------------------------------------------------------

\begin{figure}[h]
\bce\begin{picture}(100,65)
%\put(0,0){\framebox(100,65){}}
\put(0,5){\mbox{\epsfig{file=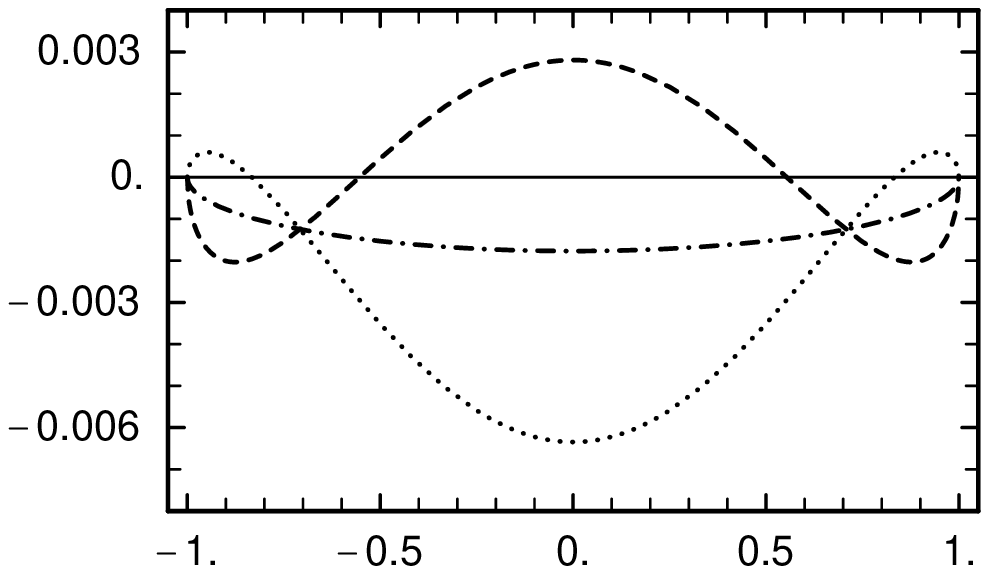,height=60mm}}}
\put(53,0){\mbox{$\cos\t_{\st}$}}
\put(-1,18){\makebox(0,0)[br]{{\rotatebox{90}{$\d\cth_{\st}([i])-
                                               \d\cth_{\st}([15])$}}}}
\put(72,55){\mbox{[10]}}
\put(76,25){\mbox{[16]}}
\put(55,34){\mbox{[17]}}
\end{picture}\ece
\caption{Differences in $\d\cth_{\st}$ between \cite{karlsruhe,beenakker3,sola} 
   and our scheme \cite{helmut} as a function of $\cos\t_{\st}$, 
   for $\mst{1}=250\gev$, $\mst{2}=600\gev$, and $\msg=600\gev$. 
   For $\d\cth_{\st}(Q^2)([16])$ we have taken $Q^2=\mst{2}^2$.}
\label{fig:deltatheta}
\end{figure}

%------------------------------------------------------------------------

\begin{figure}[h]
\bce\begin{picture}(95,65)
\put(0,5){\mbox{\epsfig{file=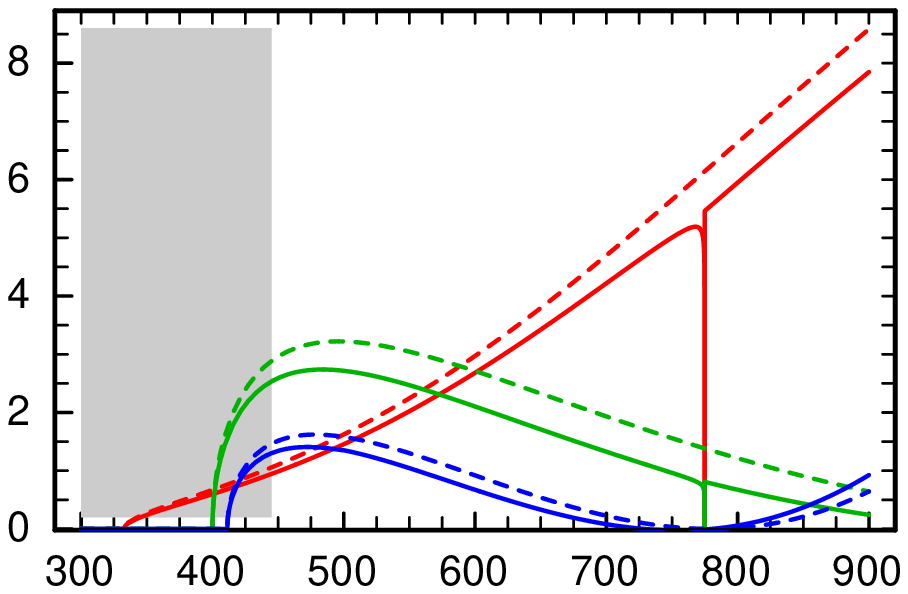,height=60mm}}}
%\put(0,0){\framebox(95,65){}}
\put(44,0){\mbox{$\mst{2}~[{\rm GeV}]$}}
\put(0,26){\makebox(0,0)[br]{{\rotatebox{90}{$\G\,(\st_2)~[{\rm GeV}]$}}}}
\put(67,50){\mbox{$\st_1 h^0$}}
\put(37,15){\mbox{$\st_1 H^0$}}
\put(35,34){\mbox{$\st_1 A^0$}}
\put(13,58){\mbox{\tiny $m_{h^0}<90\gev$}}
\end{picture}\ece
\caption{Tree--level (dashed lines) and $\Oas$ SUSY--QCD corrected 
(solid lines) decay widths of $\st_2 \to \st_1 + (h^0\!,\: H^0\!,\: A^0)$ 
as a function of $\mst{2}$, for $\mst{1}=250\gev$, $\cos\t_{\st}=0.26$, 
$\mu=550\gev$, $\tan\beta=3$, $m_A=150\gev$, and $\msg=600\gev$. 
The grey area is excluded by the bound $m_{h^0}>90\gev$.}
\label{fig:mst2}
\end{figure}

%------------------------------------------------------------------------

\begin{figure}[h]
\bce\begin{picture}(95,65)
\put(0,5){\mbox{\epsfig{file=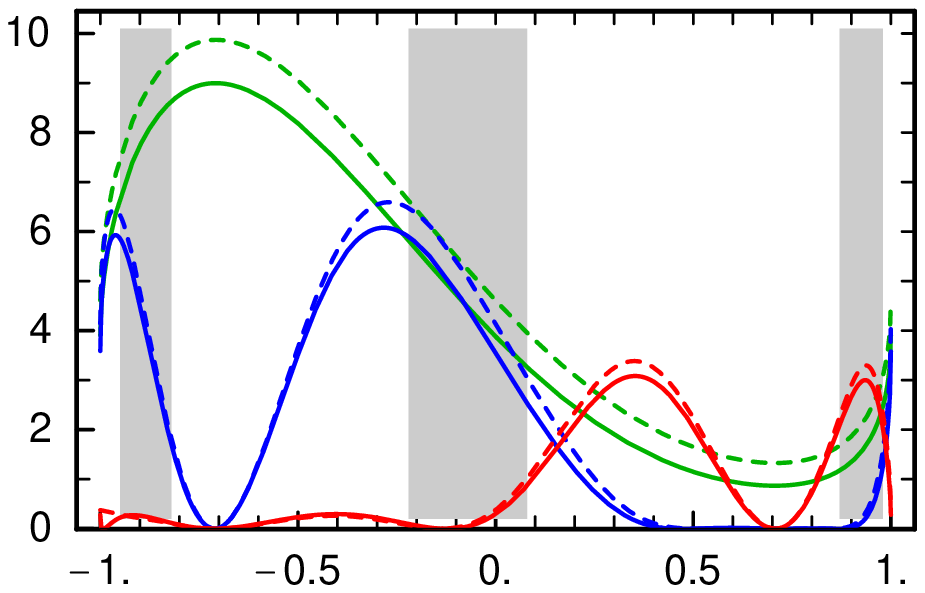,height=60mm}}}
%\put(0,0){\framebox(95,65){}}
\put(49,0){\mbox{$\cos\t_{\st}$}}
\put(0,26){\makebox(0,0)[br]{{\rotatebox{90}{$\G\,(\st_2)~[{\rm GeV}]$}}}}
\put(70,30){\mbox{$\st_1 h^0$}}
\put(35,25){\mbox{$\st_1 H^0$}}
\put(36,58){\mbox{$\st_1 A^0$}}
\put(16,60){\mbox{\tiny\bf (a)}}
\put(53,60){\mbox{\tiny\bf (b)}}
\put(90,60){\mbox{\tiny\bf (c)}}
\end{picture}\ece
\caption{Tree--level (dashed lines) and $\Oas$ SUSY--QCD corrected 
(solid lines) decay widths of $\st_2 \to \st_1 + (h^0\!,\: H^0\!,\: A^0)$ 
as a function of $\cos\t_{\st}$, for $\mst{1}=250\gev$, $\mst{2}=600\gev$, 
$\mu=550\gev$, $\tan\beta=3$, $m_A=150\gev$, and $\msg=600\gev$.
The grey areas are excluded by the constraints given in the text: 
$\d\rho(\st \mbox{--} \sb) > 0.0012$ in (a),
$m_{h^0}<90\gev$ in (b), and unstable vacuum in (c).}
\label{fig:costh}
\end{figure}

%------------------------------------------------------------------------

\begin{figure}[h]
\bce\begin{picture}(95,65)
\put(0,5){\mbox{\epsfig{file=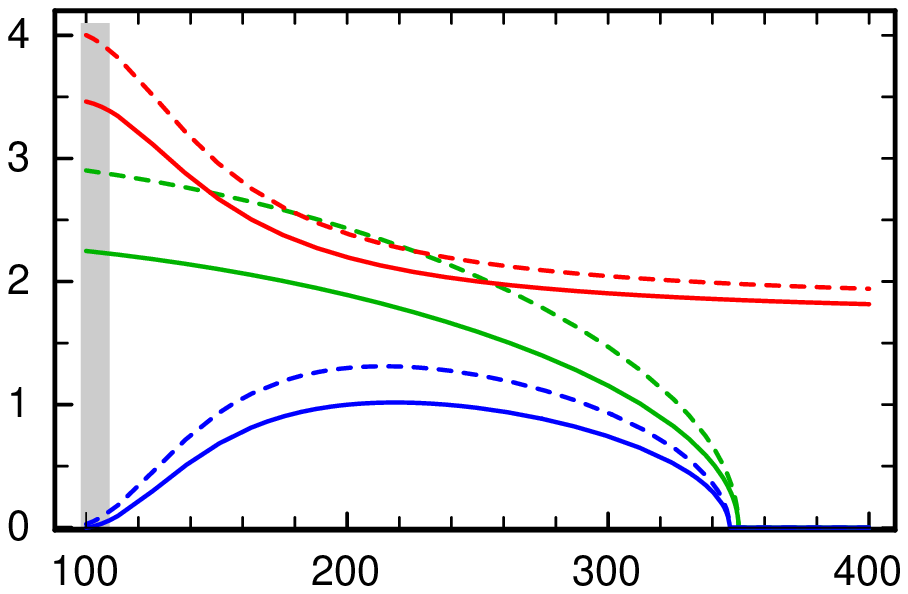,height=60mm}}}
%\put(0,0){\framebox(95,65){}}
\put(43,0){\mbox{$m_A~[{\rm GeV}]$}}
\put(0,26){\makebox(0,0)[br]{{\rotatebox{90}{$\G\,(\st_2)~[{\rm GeV}]$}}}}
\put(24,54){\mbox{$\st_1 h^0$}}
\put(17,25){\mbox{$\st_1 H^0$}}
\put(74,26){\mbox{$\st_1 A^0$}}
\end{picture}\ece
\caption{Tree--level (dashed lines) and $\Oas$ SUSY--QCD corrected 
(solid lines) decay widths of $\st_2 \to \st_1 + (h^0\!,\: H^0\!,\: A^0)$ 
as a function of $m_A$, for $\mst{1}=250\gev$, $\mst{2}=600\gev$, 
$\cos\t_{\st}=0.26$, $\mu=550\gev$, $\tan\beta=3$, and $\msg=600\gev$. 
The grey area is excluded by the bound $m_{h^0}>90\gev$.}
\label{fig:mA}
\end{figure}

%------------------------------------------------------------------------

\begin{figure}[h]
\bce\begin{picture}(95,65)
\put(0,5){\mbox{\epsfig{file=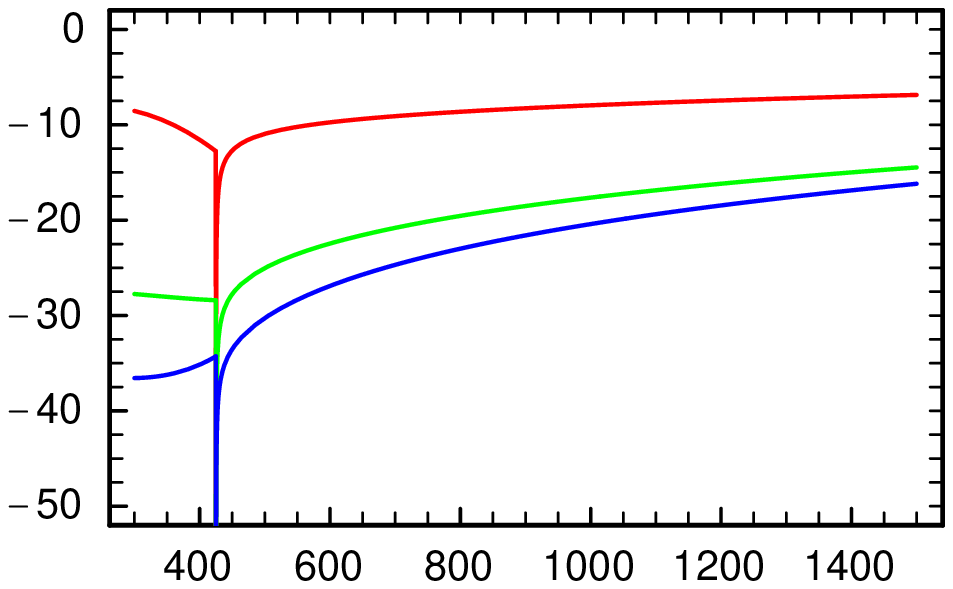,height=60mm}}}
%\put(0,0){\framebox(95,65){}}
\put(47,0){\mbox{$\msg~[{\rm GeV}]$}}
\put(0,24){\makebox(0,0)[br]{{\rotatebox{90}{$\d\G/\G^0\,(\st_2)~[\%]$}}}}
\put(27,54){\mbox{$\st_1 h^0$}}
\put(52,36){\mbox{$\st_1 H^0$}}
\put(39,45){\mbox{$\st_1 A^0$}}
\end{picture}\ece
\caption{$\Oas$ SUSY--QCD corrections (in \%) to the widths of  
$\st_2 \to \st_1 + (h^0\!,\: H^0\!,\: A^0)$ 
as a function of $\msg$, for $\mst{1}=250\gev$, $\mst{2}=600\gev$, 
$\cos\t_{\st}=0.26$, $\mu=550\gev$, $\tan\beta=3$, and $m_A=150\gev$.}
\label{fig:msg}
\end{figure}

%------------------------------------------------------------------------

\begin{figure}[h]
\bce\begin{picture}(95,65)
\put(0,5){\mbox{\epsfig{file=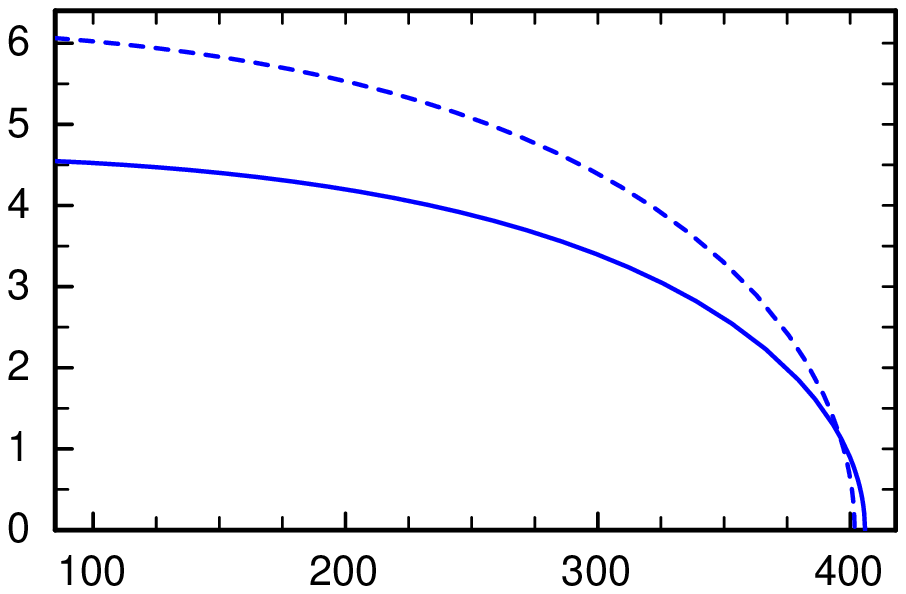,height=60mm}}}
\put(18,18){\mbox{\epsfig{file=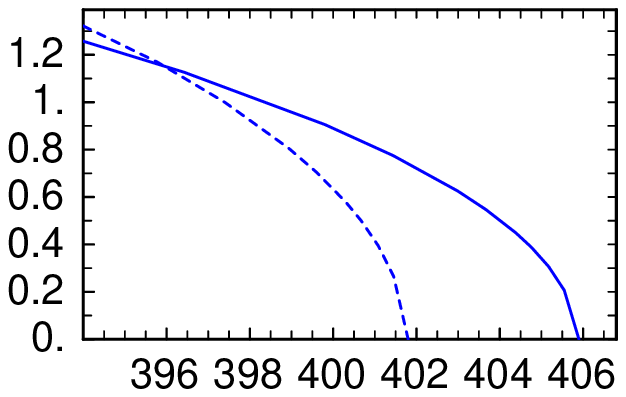,height=17mm}}}
%\put(0,0){\framebox(95,65){}}
\put(44,0){\mbox{$\mst{1}~[{\rm GeV}]$}}
\put(0,28){\makebox(0,0)[br]{{\rotatebox{90}{$\G\,(\sb_1)~[{\rm GeV}]$}}}}
\end{picture}\ece
\caption{Tree--level (dashed line) and $\Oas$ SUSY--QCD corrected 
(solid line) decay widths of $\sb_1 \to \st_1 H^-$ 
as a function of $\mst{1}$, for $\mst{2}=600\gev$, $\cos\t_{\st}=0.26$, 
$\mu=550\gev$, $\tan\beta=3$, $m_A=150\gev$, and $\msg=600\gev$. 
The insert zooms on the different thresholds at tree-- and 
one--loop level.}
\label{fig:sbmst1}
\end{figure}

%------------------------------------------------------------------------

\begin{figure}[h]
\bce\begin{picture}(95,65)
\put(0,5){\mbox{\epsfig{file=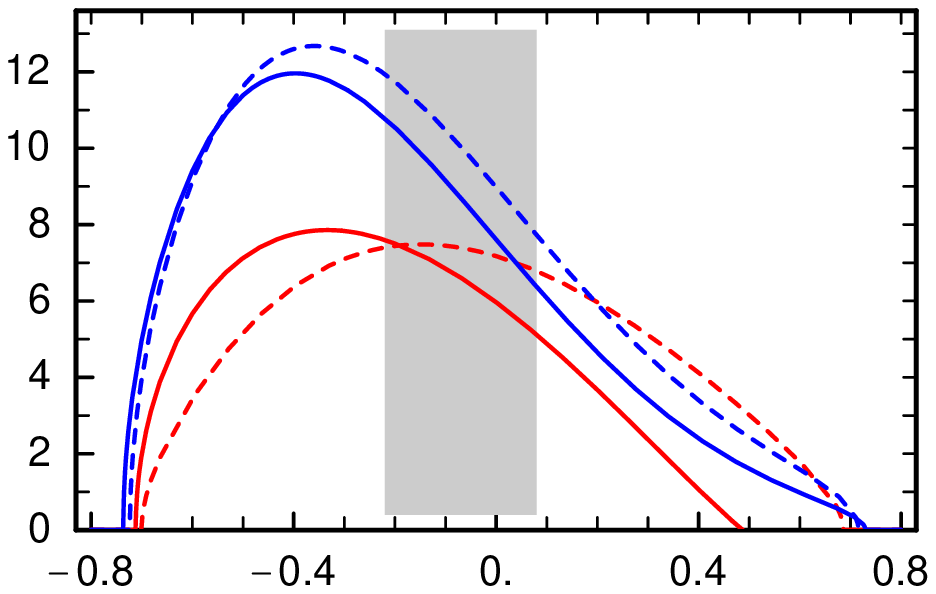,height=60mm}}}
%\put(0,0){\framebox(95,65){}}
\put(49,0){\mbox{$\cos\t_{\st}$}}
\put(0,28){\makebox(0,0)[br]{{\rotatebox{90}{$\G\,(\sb_1)~[{\rm GeV}]$}}}}
\put(27,50){\mbox{\scriptsize$\tan\b=3$}}
\put(25,24){\mbox{\scriptsize$\tan\b=10$}}
\end{picture}\ece
\caption{Tree--level (dashed lines) and $\Oas$ SUSY--QCD corrected 
(solid lines) decay widths of $\sb_1 \to \st_1 H^-$ 
as a function of $\cos\t_{\st}$, for $\mst{1}=250\gev$, $\mst{2}=600\gev$, 
$\mu=550\gev$, $m_A=150\gev$, $\msg=600\gev$, and $\tan\beta=3,\,10$. 
The grey area is excluded for $\tan\b=3$ by the bound $m_{h^0}>90\gev$.}
\label{fig:sbcosth}
\end{figure}

%------------------------------------------------------------------------

\end{document}